\documentclass[submission,copyright,creativecommons]{eptcs}
\providecommand{\event}{SecCo 2010} % Name of the event you are submitting to
\usepackage{breakurl}             % Not needed if you use pdflatex only.

\usepackage[latin1]{inputenc}

\title{Covert channel detection using Information Theory}
\author{Loïc Hélouët~~~~~~~~~ Aline Roumy
\institute{INRIA Rennes, Campus de Beaulieu, 35042 Rennes Cedex, France}
\email{loic.helouet@irisa.fr~~~~~~~~~aline.roumy@inria.fr}
}

\def\titlerunning{Covert channel detection using Information Theory}
\def\authorrunning{L.Hélouët, A. Roumy}

\usepackage{amssymb}
\usepackage{amsmath}
\usepackage{amscd}
\usepackage{enumerate}
\usepackage{epsfig}

\newtheorem{definition}{\textbf{Definition}}
\newtheorem{theorem}{\textbf{Theorem}}

\newcommand{\beq}{\begin{eqnarray}}
\newcommand{\eeq}{\end{eqnarray}}

\begin{document}
\maketitle

\begin{abstract}
This paper presents an information theory based detection framework for covert channels. We first show that the usual notion of interference does not characterize the notion of deliberate information flow of covert channels. We then show that even an enhanced notion of ``iterated multivalued interference'' can not capture flows with capacity lower than one bit of information per channel use. We then characterize and compute the capacity of covert channels that use control flows for a class of systems.
\end{abstract}

\vspace{-2\medskipamount}
\section{Introduction}
\vspace{-\medskipamount}

The term covert channel was first introduced by Lampson~\cite{Lampson73}, and designates an information flow that violates a system's security policy. In a system, this policy can define who is allowed to communicate with whom, through which channels, and forbid all exchanges other than these legitimate ones. Security policies can also define filtering or billing policies when legal channels are used, and which exchanges should be observed and recorded. They can be implemented by system monitors, that ensure that unauthorized communications do not occur, and record some events of the system. Within this context, a {\em covert channel} is a perverted use of a system by two legal users. These users have access to system's functionalities, but use them in a way that bypasses the security policy (for instance to create a communication channel between two users that are not allowed to communicate, or to pass information between authorized users without paying for it, etc.). One usual assumption is that both corrupted users know perfectly the system, and have agreed on a particular use of the functionalities to encode and decode information. 

Unsurprisingly, this preoccupation for covert information flows appeared in the 70's, with a particular attention paid to information systems. The fear during this period was that an agent with high level accreditation would read classified information, and send them discretely to another agent with low accreditation. This covert channel problem also has an economic interpretation: covert flows can be used to establish free communications over paying services. Nowadays, with the increase of online transactions and personal computers, the problem seems more individual: the threat is that a Trojan horse can communicate personal information (agenda, credit card numbers,...) to a third party via covert channels that would bypass all protections of the computer (firewalls, anti viruses,...).

Many security recommendations~\cite{Common99,NCSC93} consider covert channels in their lists of threats, and ask for the application of reproducible methods to characterize channels, evaluate their capacity, and depending on the severity of the threat, to close or lower the information leak. Of course, ``reproducible methods'' advocates for the use of formal models and formal techniques. Many model-based methods have been proposed, such as shared matrices, non-interference checking, etc. Note however that it is commonly agreed that one particular technique can not capture all kinds of information leaks. The first formal model allowing the automation of security leaks discovery is the well-known Bell \& La Padula model introduced in~\cite{Bell73a,Bell73b}. It can be modeled as a matrix~\cite{Kemmerer83} defining accesses of agents to objects in the system, and a security leak occurs if the transitive closure of the matrix contains a forbidden access. Since the 80's, information leak is mainly considered through the notion of {\em interference}~\cite{Goguen82}, that characterizes information leak from a high level (confidential) part of the system to low level (public) part. At first sight, this looks very similar to the definition of covert flows: one can search for interference from $u$ to $v$ by declaring as confidential all actions of $u$, and public all actions of $v$. 
%
% However, we will show that interference and covert channels are distinct notions of information leaks. The first reason is that interference does not capture the deliberate nature of leaks that is inherent to covert channels. Furthermore, as long as interference is defined as a system equivalence, it misses channels where less than one bit of information per use of the covert channel is leaked. 
%
%INSERT *1
%
However, we will show that interference and covert flows are orthogonal notions of information leakage. A first difference is that covert flows are situations in which two agents cooperate to allow transfer of information, while an interference means that some classified or confined information can be recovered by an agent from its observations of the running system, without collaboration. A second difference is that corrupted users of a covert flow must be able to transfer any message of arbitrary size in a bounded time (this is called the ``small message criterion''~\cite{Moskowitz94}), while leaking a single bit of information in a run, or the same information an arbitrary number of times is sufficient for a system to be interferent. 
An immediate idea to extend interference is to consider a notion of iterated, deliberate and multivalued interference. We will detail this possibility in the paper, and show that such notion still misses some obvious covert flows. We hence propose a new characterization for covert flows in systems modeled as transition systems. This characterization considers that a covert channel exists between two users $u$,$v$ if $u$ and $v$ can use the system to simulate a memoryless discrete channel with state of capacity greater than $0$. 

This paper is organized as follows: section~\ref{section_non_interference} introduces the notion of interference and the formal material that is used in the paper. Section~\ref{section_NI_limits} shows some differences between the notions of interference and covert channels. The notion of communication channel used in the paper needs some elements of information theory, that are introduced in Section~\ref{section_information_theory}. Section~\ref{section_first_tries} shows that former trials that extend the notion of interference either through a notion of ``iterated'' interference , or via the quantification of common knowledge of processes fail to characterize covert channels. Section~\ref{section_example} is an easy covert channel example that illustrates the characterization of covert channels proposed in section~\ref{section_characterization}. Section~\ref{section_conclusion} discusses some technical choices, and concludes. Some details omitted in the paper can be found in an extended version at {\tt www.irisa.fr/distribcom/Personal\_Pages/helouet/Papers/Secco2010\_extended.ps}.

\vspace{-2\medskipamount}
\section{Non-interference}
\label{section_non_interference}
\vspace{-\medskipamount}

The term {\em non-interference} was first introduced in~\cite{Goguen82}. In the original definition, an agent $u$ interferes with an agent $v$ in a system $S$ iff ``what $u$ does can affect what $v$ can observe or do''. The proposed model on which interference is checked is a kind of transition system, in which moves from one state to another are performed by one agent, and where each agent has in addition some ``capacities'' that allow him to test inputs/outputs to the system or observe the values of some variables in each state of the system. Within this context, we can see $S$ as a system composed of several agents and subsystems, i.e.  $S=u\mid p_1 \mid \dots \mid p_k \mid v$, and $u$ does not interfere with $v$ in system $S$ iff the system behaves similarly from $v$'s point of view, independently from the fact that $u$ performs some actions or not. This can be written formally as: $ \Pi_v(u\mid p_1 \mid \dots \mid p_k \mid v) \sim \Pi_v(p_1 \mid \dots \mid p_k \mid v) $, 
where $\Pi_v(S)$, the projection of $S$ on process $v$ represents what $v$ can observe from $S$, and $\sim$ is an equivalence relation between systems. 

This latter definition of non-interference is very generic, as we have not precised the nature of $S, \Pi,$ or $\sim$. One may immediately notice that non-interference is not uniquely defined for a kind of system, and depends on the considered equivalence, which in some sense captures the discriminating power of an observer of the system. The choice of a given kind of model and of an equivalence between models (trace equivalence, bisimulation, testing equivalence,...) allows the definition of a large variety of non-interferences. 
To be able to decide if there exists an interference between $u$ and $v$, non-interference must rely on a decidable equivalence relation for the models used to represent the behaviors of the system. For instance, if $S$ is modeled by communicating automata, and $\sim$ is trace equivalence, then non-interference is undecidable. 
%
%We will get back to non-interference later in this report, and will show that it characterizes the fact that a system leaks some confidential information, but that it is not always sufficient to characterize deliberate information flows, and in particular covert channels. 
%Non-interference has been proposed for a variety of formal models. 
In the rest of the paper, we will use 
%a widely accepted model, namely 
{\em transition systems} to represent distributed systems behaviors. 

\vspace{-\medskipamount}
\begin{definition}
A {\em transition system} is a tuple $S=(Q,\longrightarrow, \Sigma,q_0)$ where
$Q$ is a finite set of states, $q_0$ is the initial state of the system, $\longrightarrow \subseteq Q\times \Sigma\times Q$ is a transition relation, $\Sigma$ is an alphabet of actions. We will furthermore consider the unobservable action $\tau \not\in \Sigma$ such that for all  $a\in \Sigma$, $\tau.a=a.\tau=a$.  
Transition systems define the behaviors of a set of processes $\mathcal P$. Each action in $\Sigma$ is executed by a single process, and observed by several processes. This is modeled by two functions $Ex: \Sigma \longrightarrow \mathcal P $ and $Obs: \Sigma \longrightarrow 2^{\mathcal P}$.
\end{definition}

\vspace{-\medskipamount}
\noindent A {\em path} in a transition system $S$ is a sequence of transitions 
%of the form 
$\rho=(q_1,\sigma_1,q_2)(q_2,\sigma_1,q_3) \dots
(q_{k-1},\sigma_{k-1},q_k)$. We will also write $\rho= q_1 \stackrel{\sigma_1}\longrightarrow q_2 \stackrel{\sigma_2}\longrightarrow q_3 \dots
q_{k-1}\stackrel{\sigma_{k-1}}\longrightarrow q_k$, and denote by $Path(x,y)$ the set of paths starting in $x$ and ending in $y$.

\begin{definition}
The {\em language} of a transition system $S$ is the set of words  $\mathcal
L(S) \subseteq\Sigma^*$ such that for all $w=\sigma_1.\sigma_2\dots \sigma_k\in \mathcal L(S)$ there exists a path $q_0 \stackrel{\sigma_1}\longrightarrow q_1
\stackrel{\sigma_2}\longrightarrow \dots \stackrel{\sigma_k}\longrightarrow
q_k$ starting in the initial state of $S$. We will say that two transition systems $S$ and $S'$ are {\em equivalent} (denoted by  $S\sim S'$) iff $\mathcal L(S)=\mathcal L(S')$.
\end{definition}
\vspace{-\medskipamount}

As we want to consider systems in which some processes (for instance the environment) behave non-deterministically, we will attach to the firing of transitions from a given state a discrete probabilistic distribution. We then associate to a transition system $S$ a probability function $P_S: Q\times \Sigma\times Q \longrightarrow \mathbb R$, with the constraint that $\forall q\in Q, \sum_{q'\in Q,a\in \Sigma} P_S(q,a,q')=1$. To simplify notations, $P_S$ is only partially defined, and we assume a uniform distribution on outgoing transitions from each state $q$ for which $P_S$ is not defined. Function $P_S$ also allows for the probabilization of paths and words. For $\rho=q_0 \stackrel{\sigma_1}\longrightarrow q_1
\stackrel{\sigma_2}\longrightarrow \dots \stackrel{\sigma_k}\longrightarrow
q_k$, we have $P_S(\rho)=P_S(q_0,\sigma_1,q_1).P_S(q_1,\sigma_2,q_2) \dots P_S(q_{k-1},\sigma_k,q_k)$.

\vspace{-\medskipamount}
\begin{definition}
The {\em projection} of a transition system $S=(Q,\longrightarrow, \Sigma,q_0)$
over an alphabet $X\subseteq \Sigma$ is the system $\Pi_X(S)=(Q,\longrightarrow',
\Sigma,q_0)$, such that $\longrightarrow'=\{(q,a,q')\mid a\in X\} \cup \{
(q,\tau,q') \mid \exists (q,a,q') \wedge a\not\in X\}$. The {\em restriction} of $S$ to $X$ is the system $S_{\setminus X}=(Q'',\longrightarrow'',\Sigma,q_0)$, where 
$\longrightarrow''= \{(q,a,q')\mid a\in X\}$ and $Q''$ is the restriction of $Q$
to states that remain accessible via $\longrightarrow''$.
\end{definition}
\vspace{-\medskipamount}

Projections and restrictions can be used to focus on a specific aspect of a system:  $\Pi_{Obs^{-1}(u)}(S)$ defines what process $u$ observes from 
$S$. $S_{\setminus Ex^{-1}(u)}$ defines allowed behaviors of $S$ when process $u$ does not perform any action. To simplify notations, we will write  
  $\Pi_u(S)=\Pi_{Obs^{-1}(u)}(S)$ and  $S_{\setminus u} =S_{\setminus Ex^{-1}(u)}$. Of course, projections extend to words and languages by defining for every  $X\subseteq \Sigma$, every $a\in \Sigma$, and every word 
$w\in \Sigma^*$ the projection as  $\Pi_X(a.w)=\Pi(w)$ if  $a\not\in X$, and 
$\Pi_\Sigma(a.w)=a.\Pi_X(w)$ otherwise. 
%Using these notions of restriction and projections, we can now formalize a definition of interference. 

\vspace{-\medskipamount}
\begin{definition}
\label{def_interference}
User $u$ {\em interferes} with user $v$ in system  $S$ if and only if $\Pi_v(S)\not\sim \Pi_v(S_{\setminus u})$
\end{definition}
\vspace{-\medskipamount}

More intuitively, definition \ref{def_interference} says that what user $v$ sees from system $S$  (i.e. $\Pi_v(.)$) changes when process $u$ is allowed to do some actions or not (i.e. if $v$ can  distinguish if it observes
$S$ or $S_{\setminus u}$). This definition is only one among many definitions of interference. It is usually called SNNI (Strong Non-deterministic Non Interference) in the literature. 
A similar notion called BSNNI (Bisimulation based SNNI) exists where $\sim$ is replaced by a bisimulation relation. 
% One can notice that modifying either the considered model, the observation capacities of processes (the way $\Pi_v(.)$ is defined) or the equivalence relation $\sim$ yields different definitions of interference. Hence, there exists a huge number of interferences. 
We refer interested reader to~\cite{Focardi00b}, which defines, compares and classifies several interferences for systems described with process algebra. Another interesting state of the art 
%on interference 
can also be found in~\cite{Sabelfeld03}. 
% At this point, one should notice that to be effective, the equivalence relation must be decidable on the considered class of models (one can not for instance test BSNNI for Petri nets~\cite{Esparza94} or CFMS).
In the rest of the paper, we will focus on SNNI, but keeping in mind that the differences highlighted in sections~\ref{section_NI_limits} and~\ref{section_first_tries} hold for interference in general.
% Interference has also been defined through typing approaches~\cite{Boudol01,Volpano96}, but we will not detail this approach in this document. 
%
%
%INSERT 2
%
Note also from definition~\ref{def_interference} that during his observation of a system $S$, an agent $v$ may observe sequences of actions that provide him with some information on $u$'s behavior, but that no cooperation from $u$ is a priori needed to get this information. 

\vspace{-2\medskipamount}
\section{First differences between interference and covert flows}
\label{section_NI_limits}
\vspace{-\medskipamount}

\begin{figure}[htbp]
\begin{center}

\epsfig{figure=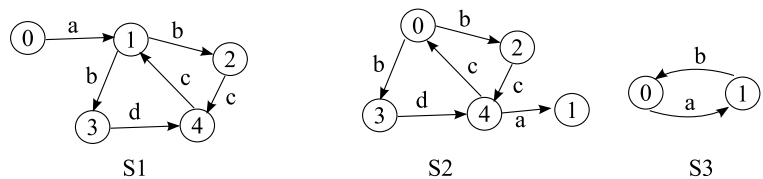, width=10cm}
\vspace{-2\medskipamount}
\caption{Examples of interferent systems}
\label{interferent_examples}
\end{center}
\vspace{-2\medskipamount}
\end{figure}

An interference in a distributed system means that a process of the system (or a user) can obtain some confidential information on values of variables, or about other users behaviors through its observations. Several papers consider that  covert channels are a sub-case of interference. In this section, we will show that interference captures a notion of {\em information leak}, but does not necessarily characterize {\em deliberate information flows} of arbitrary size. Let us consider the examples of Figure~\ref{interferent_examples}, that depict the behavior of systems involving two users $u$ and $v$, and where action $a$ is executed and observed by $u$, and actions $\{b,c,d\}$ are observed and executed by $v$. For these three transition systems, the initial state is state $0$. In $S1$, user $u$ can perform action $a$, and then user $v$ can execute any prefix of $(bcc+bdc)*$. The projection $\Pi_v({\mathcal L}(S1))$ is the set of prefixes of
$(bcc+bdc)^*$, and the projection $\Pi_v({\mathcal L}(S1_{\setminus u}))$ is 
the empty word $\epsilon$. Hence, from definition~\ref{def_interference}, processes $u$ interferes with $v$ in $S1$. Note however that this interference is due to a single transition, which can be fired only once in each execution of the system. System $S2$ depicts the converse situation: user $v$ can execute any sequence of actions in  $(bcc+bdc)^*(bc+bd)$ before user $u$ executes action $a$, hence $v$ interferes with $u$ in $S2$. However, when $u$ executes $a$, it is impossible for him to detect which sequence of actions of $v$ occurred before $a$. Furthermore, after executing $a$, the system remains deadlocked in state $1$. Transition systems $S1$ and $S2$ are example of interferent specification where interference occurs {\em only once}, and which can not be used to transmit a message of {\em arbitrary size}, as usually expected in covert channels.

Considering covert channels as a sub-case of interference implicitly supposes that the detected interferences can be repeated an arbitrary number of time, to transmit a message of arbitrary size. This means that when a system always reaches a sink state after interfering, it is then set back to its initial state. We think that this interpretation does not hold for most systems (for instance when sink states represent faulty deadlocked states), and that system resets should be explicitly modeled in the specification if they can occur. Note also that covert channels are supposed to be as discreet as possible, and that causing a fault that needs resetting a system does not really comply with this assumption.

The last system $S3$ in Figure~\ref{interferent_examples} contains interferences from $u$ to $v$ and from $v$ to $u$. These interferences can be repeated an arbitrary number of times, which would a priori allow for the encoding of a message of arbitrary size. However, user $v$ observes a sequence of $b$'s of arbitrary size, without knowing whether the covert message has been completely transmitted or not. This situation hence does not allow an encoding of a covert message. Yet, there is a possibility to pass some information from $u$ to $v$ (or conversely) if time can be used as a vector of information, that is if $u$ and $v$ can measure the time elapsed between two $a$'s or $b$'s. Processes $u$ and $v$ can for instance agree that if action $a$ is fired within a short time interval after it is enabled, it means bit $0$, and if the same action is delayed, it means bit $1$. Covert channels that use time measurement to pass information are frequently called {\em timing channels}. 
%However, using time to pass information supposes that the system preserves the delays that are chosen by the sender in the covert channel, and that the receiving process has an accurate measure of time. 
We will not address timing channels in this paper, and focus on covert channels that use control flows of systems to transfer information. 
%
%INSERT *3
%
Note however that time elapsing and measurement can respectively be seen as the inputs and outputs of an untimed channel, and that the characterization of section~\ref{section_characterization} may still work in the timed cases. 

\vspace{-2\medskipamount}
\section{Information Theory}
\label{section_information_theory}
\vspace{-\medskipamount}

%% START = modified section By Aline
The characterization proposed in section~\ref{section_characterization} for covert flows shows that a pair of users can simulate a {\em memoryless communication channel}, an usual notion of coding theory.  We hence recall some elements of information theory that are needed in the rest of the paper. Interested readers are referred to~\cite{CoverThomas} for a reference book on information theory. 

A {\em discrete random variable} $X$ takes values in a discrete alphabet $\mathcal X$ and is characterized by its probability mass function $P(X=x)$ that gives the probability that $X$ is exactly equal to some value $x\in \mathcal X$. In the sequel, we will denote by $p(x)$ this probability. Consider for example the reduced card set of Figure~\ref{jeu_de_cartes}-a:

\begin{figure}[htbp]
\vspace{-9\medskipamount}
\begin{center}
\epsfig{figure=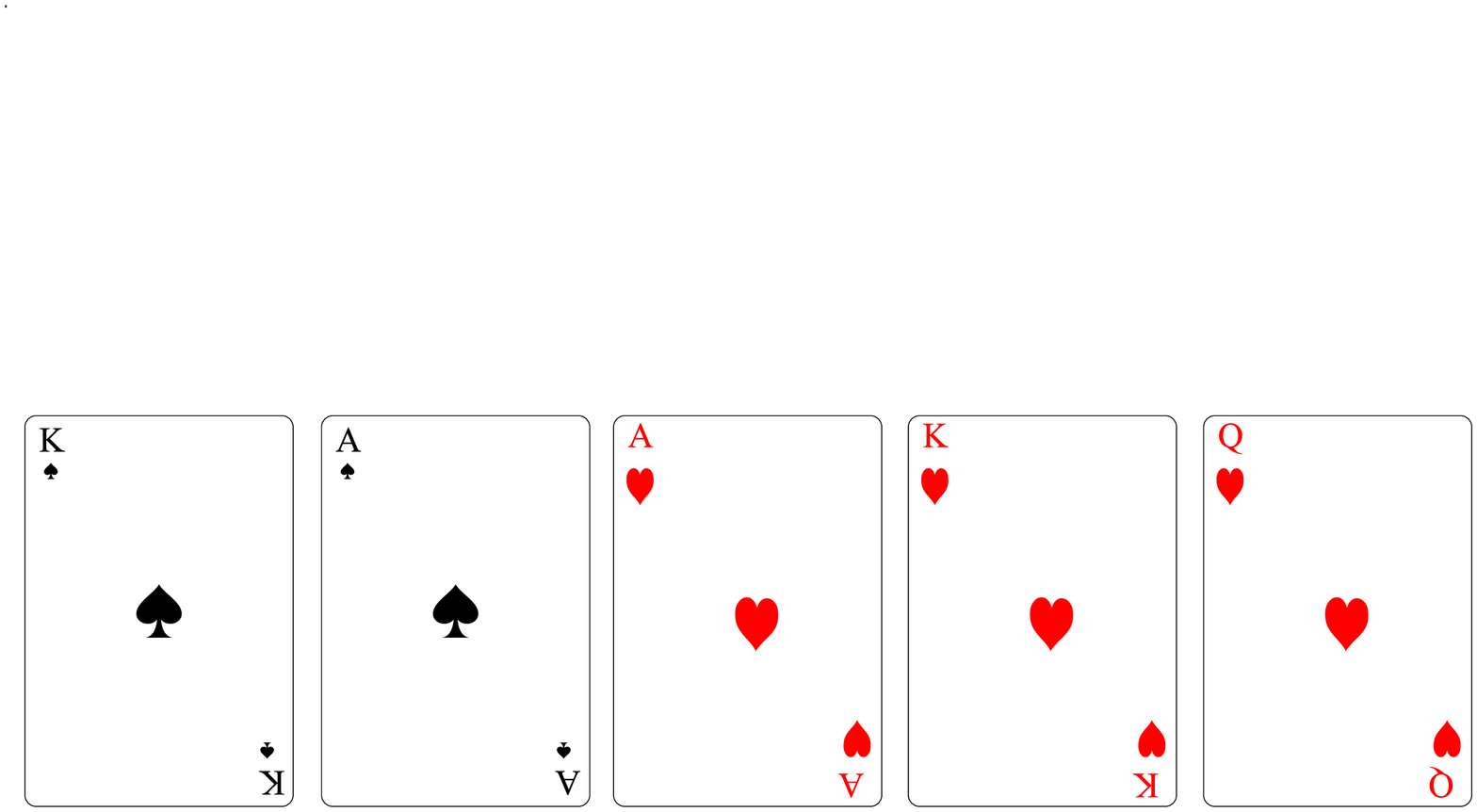,width=8cm}\hspace{1cm}\epsfig{figure=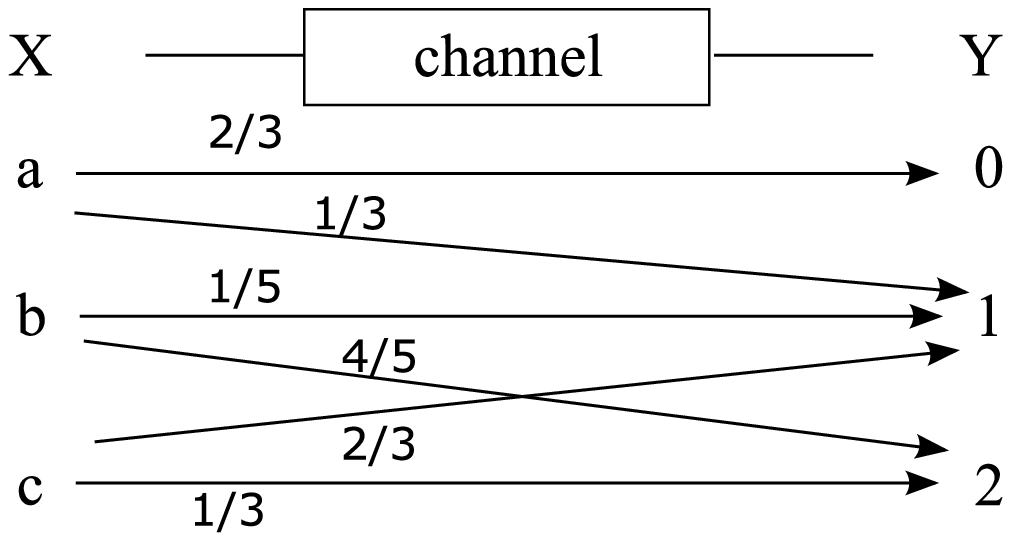, width=5cm}
\vspace{-\medskipamount}
\caption{~~~~~~a) A reduced card set~~~~~~~~~ b) A discrete channel model}
\label{jeu_de_cartes}
\end{center}
\vspace{-3\medskipamount}
\end{figure}

One can consider as random variable $Card$ the complete name of a card, that is a pair $(value, color)$ in a set $\{ (King, \spadesuit), (Ace, \spadesuit), (Ace, \heartsuit), (King, \heartsuit), (Queen, \heartsuit)\}$, and associate probability $1/5$ to each value. Now, if we only consider the color of the card, we define a random variable (say $Color$) with value set $\{ \spadesuit, \heartsuit\}$, and associated probabilities $2/5$ and $3/5$. We can also define another random variable $Value$, over a domain $\{ King, Ace, Queen\}$, with associated probabilities $2/5,2/5,1/5$. The random experience can be repeated, and we will also consider sequences of random variables $X_1.X_2\dots X_n$, denoting $n$ consecutive choices of a value for $X$. In the sequel, $x_n$ denotes the $n^{th}$ value taken by the variable $X_n$ and we will write $X^n$ (resp. $x^n$) instead of $X_1\dots X_n$ (resp. $x_1\dots x_n$).
%, and denote by $x_n$ the $n^{th}$ value taken by variable $X$.

The {\em {\bf entropy}} (expressed in bits) is a measure of the uncertainty associated with a random variable, and is defined as $H(X) = -\underset{x\in \mathcal X}\Sigma p(x).\log_2 p(x)$.
In some sense, entropy measures the average number of binary questions to ask to know the value of a random variable after a random choice. Let us get back to the card set of Figure~\ref{jeu_de_cartes}-a. The entropy of variable $Card$ is $H(Cards)=-5.(1/5).\log_2(1/5)=2.32$. Now, our card set can be seen as a pair of random variables $Value,Color$. Let us randomly choose a card. Knowing the card set, we can apply the following strategy to guess the correct pair of random variables: first, discover the color of the card (this is done with a single question), and then discover its value (this can be done asking at most two questions). The average total number of question to ask with this strategy is $2.4$ and this is more efficient than enumerating all values, which leads to asking $2.8$ questions on average. Note also that knowing the color of a card provides some information on its value: if a card is a spade, then it is useless to ask whether it is a queen to discover the chosen card. This is explained by the fact that variables $Color$ and $Value$ are not independent. The quantity of information shared between two random variables is called the  {\em {\bf mutual information}}, and is defined as $I(X;Y)=H(X)+H(Y)-H(X,Y)$.

From this definition, one can show that $I(X;Y)=H(X)-H(X|Y)$, which provides a quite intuitive explanation of mutual information. The mutual information between $X$ and $Y$ is the uncertainty on $X$ minus the uncertainty on $X$ that remains when $Y$ is known. This notion of mutual information will be used later to evaluate the capacity of communication channels. This value is symmetric, so we have $I(X;Y)= H(X)-H(X|Y) = H(Y)-H(Y|X) = I(Y;X)$

In communication systems, a frequent challenge is to measure the amount of information that can be passed from a source to a destination. The communication medium can be a wire, a network, an hertzian channel, etc. Transmissions from the source to the destination are encoded and transmitted along the medium to the receiver with a fixed rate. One usually considers that the symbols that are input in the medium and those that are received at the other end, have not necessarily the same alphabets. Furthermore, the channel can be noisy, and there is no 
one-to-one correspondence between the inputs and outputs of the channel.  
This situation is modeled as a triple $(X,Y,p(y\mid x))$, where $X$ (resp. $Y$) is a random variable taking values in $\mathcal X$ (resp. $\mathcal Y$), the set of input (resp. output) values of the channel, and $p(y\mid x)$ is a conditional probability law 
%on inputs/outputs of the system, 
that associates to each input of $x\in\mathcal X$ the probability to obtain $y\in\mathcal Y$ as an output. This probability is usually referred to as the transition probability of the channel.
%
% \begin{figure}[htbp]
% \begin{center}
% \epsfig{figure=channel.eps, width=7cm}
% \caption{A discrete channel model}
% \label{canal_model}
% \end{center}
% \end{figure}
%
Communication channels are usually represented as in Figure~\ref{jeu_de_cartes}-b: the input symbols of the channel are placed on the left of the picture, the output symbols on the right, and the probability to obtain some $y\in \mathcal Y$ after sending a $x\in \mathcal X$ in the channel is depicted by an arrow from $x$ to $y$, labeled by the conditional probability $p(y\mid x)$. The channel represented on Figure~\ref{jeu_de_cartes}-b has input alphabet $\{a,b,c\}$ and output alphabet $\{0,1,2\}$. The probability to obtain $0$ as an output of the channel after sending symbol $a$ is $2/3$. 
 
The {\em capacity} of a channel is the average mutual information per use of the channel. It is defined as $C=\lim_{n\rightarrow \infty} ~~\underset{ p(x^n)}\max~~ \frac{1}{n}I(X^n;Y^n)\label{general_capacity_def}$
where $X^n$ are the $n$ possible inputs sent over the channel and $Y^n$ are the $n$ consecutive symbols received. The maximization is performed over all probability mass functions of the input sequence. 
Note that one can transfer {\em less than one bit} of information at each use of the channel. Note also that this capacity formula involves a maximization which might not be feasible in practice. When a discrete channel is {\em memoryless}, the capacity reduces to $C= \underset{ p(x)}\max ~~ I(X;Y)\label{letter_characterization}$. Such a closed form is called a ``single letter characterization'' of the channel capacity. It is simple to compute, as it only involves a maximization over a set of probability mass functions of a single discrete random variable. However, single letter characterizations do not necessarily exist for all kinds of channels.  
 Let us recall at this point that a channel is seen as a fixed rate use of a communication medium, and that every use of the channel is performed within a time period $T$. Hence, time elapsed between two consecutive uses of the channel is not considered within this setting as carrying information.

\begin{figure}[htbp]
\begin{center}
\epsfig{figure=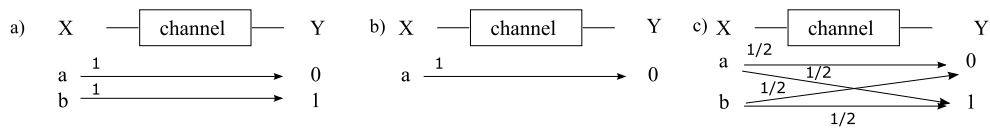, width=13cm}
\vspace{-2\medskipamount}
\caption{Some well known discrete channels}
\label{fig_various_channels}
\end{center}
\vspace{-2\medskipamount}
\end{figure}

Let us now review some useful discrete memoryless channels. First consider the {\em perfect channel}, depicted in Figure~\ref{fig_various_channels}.a). This channel is noiseless and establishes a one-to-one correspondence between inputs and outputs. Therefore its capacity is $\log_2(|\{ a,b\}|)=1$, where $|\mathcal X|$ is the size of the set $\mathcal X $ i.e. one can transfer one bit of information at each use of the channel. In a more general way, the capacity of a channel of this kind that associates a unique output to every input with probability 1 is $\log_2(|\mathcal X|)$. Similarly, there are zero-capacity channels. Consider for instance the channel of Figure~\ref{fig_various_channels}-b. The capacity of this channel is $C= - 1.\log_2(1)=0$. This is not surprising, as there is no possibility to encode two distinct values.
The capacity of the channel of Figure~\ref{fig_various_channels}-c is also zero, as $C= \underset{\ p(x)}\max ~I(X;Y) = \underset{p(x)}\max ~H(Y)-H(Y\mid X)$ and for every distribution over $\mathcal X$, $H(Y\mid X)=H(Y)$. A similar property holds for an arbitrary number of inputs and outputs for channels where $p(y\mid X=x)$ is uniform for every input $x\in \mathcal X$. 

% version actuelle
In this paper, we want to test if a transition system contains a communication channel. By definition, the possible actions taken by a process (i.e. the inputs of the channel) depend on the state of the system. Therefore, the communication channels we are looking for are {\em state channels} as defined below: 

\begin{definition}[State Channel]
 \label{def_state_channel}
A memoryless discrete channel with independent and identically-distributed (i.i.d.) {\em state} is a tuple $K=(S,p(s),\{X_s,Y_s,p(y\mid x,s) \}_{s\in \mathcal S})$, where 
$S$ is random variable defined over a set $\mathcal S=\{1,\dots, h \}$, $p(s)$ is the probability for the system to be in state $s\in \mathcal S$.
$X_s, Y_s$ are random variables respectively ranging over a set of input/output  values $\mathcal X_s$ (resp. $\mathcal Y_s$) and represent the input and output of the channel. The choice of the state is statistically independent of previous states and previous input or output letters in the channel.
$p(y\mid x,s)$ is the conditional distribution that defines the probability to get $y\in \mathcal Y_s$ as output when the channel is in state $s\in \mathcal S$ and the input $x\in \mathcal X_s$ is transmitted. 
 \end{definition}
\vspace{-\medskipamount}

This definition differs from the one given in \cite{Shannon58} since here the input alphabets depend on the state. Moreover, Shannon has noticed that some additional information is frequently available at the transmitter, such as the state of the communication channel. Knowing this information can help increasing the capacity of the channel~\cite{Shannon58}.

\begin{theorem}
\label{thm_state_channel_capacity}
Let $K=(S,p(s),\{X_s,Y_s,p(y\mid x,s) \}_{s\in \mathcal S})$ be a memoryless discrete channel with i.i.d. states defined in Definition~\ref{def_state_channel}.
The capacity of the channel $K$  with side state information known causally at the transmitter (at time $n$, the current state $S_n$ is known)
is equal to the capacity  $C= \underset{ p(t)}\max ~~ I(T;Y)$ of the memoryless channel $K'=(T,Y,r(y\mid t))$ (without side information) with the output alphabet $\mathcal Y = \cup_{s \in \mathcal S} \mathcal Y_s$ and an input alphabet 
$\mathcal T=\mathcal X_1\times \dots \times\mathcal X_{|\mathcal S|}$. 
The transition probabilities of the channel are given by:\\
$r(y\mid t)=\underset{s\in \mathcal S}\sum p(s).p(y\mid t(s),s)$, where $t(s)$ stands for the $s^{th}$ component in the vector $t$.
\end{theorem}
\vspace{-\medskipamount}
The proof follows the same lines as in~\cite{Shannon58}. The difference is that in ~\cite{Shannon58} each $t$ is a particular function from the state alphabet $\mathcal S$ to the input alphabet of the original channel $\mathcal X$ whereas here $t$ is a vector that spans the set $\mathcal X_1\times \dots \times\mathcal X_{|\mathcal S|}$.

\begin{figure}[htbp]
\vspace{-3\medskipamount}
\begin{center}
\epsfig{figure=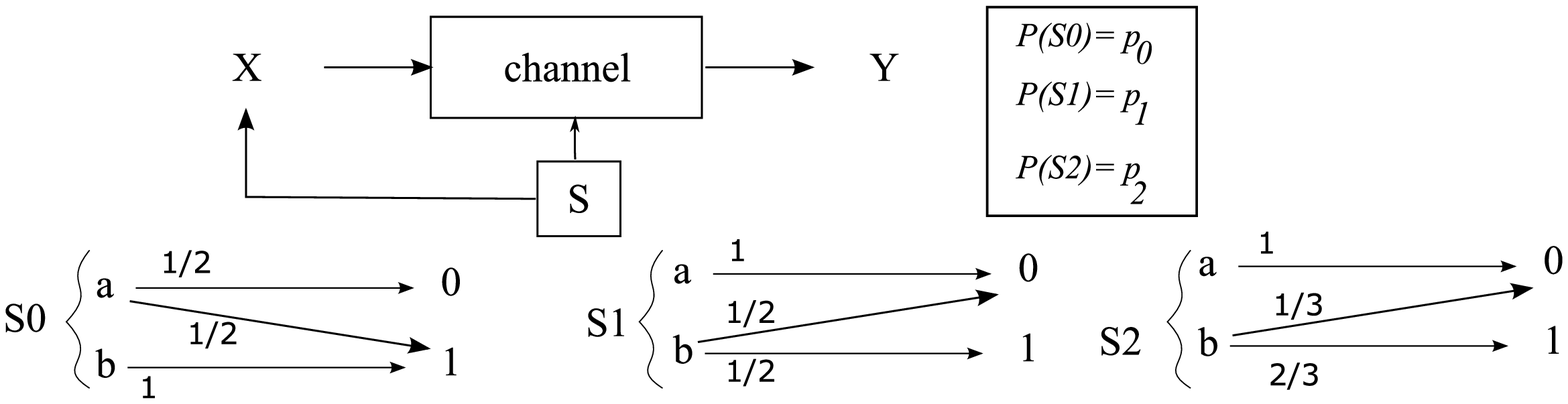, width=10cm}
\vspace{-2\medskipamount}
\caption{A memoryless discrete channel with side state information}
\label{fig_state_channel}
\end{center}
\vspace{-2\medskipamount}
\end{figure}

\vspace{-3\medskipamount}
\section{More differences between interference and covert flows}
\label{section_first_tries}
\vspace{-\medskipamount}

As shown in previous sections, interference occurs when a process can learn some hidden information from his observation of the system, but does not always characterize covert flows of information. In section~\ref{section_characterization}, we show a characterization that highlights presence of a covert flow while measuring its capacity. It is important to note two important facts on covert channels: first, efficient coding techniques do not impose to send at least one bit of information per use of the channel. Second, the notion of mutual information between processes behaviors (i.e observable sequences of actions) does not necessarily characterize a {\em deliberate} information flow. Hence, one must be able to differentiate between coincidence and covert flows. In the sequel, we first show that the intuitive extension of interference to a notion of {\em iterated interference} with several distinguishable values (as proposed in~\cite{hzd04}) fails in general to capture covert flows with capacity $<1$. We then show that mutual information measurement between processes (as proposed in~\cite{Millen87}) does not capture either the idea of a covert protocol between users of a covert channel, and may then consider as a covert flow a situation where two processes observe the same phenomena. 

\vspace{-2\medskipamount}
\subsection{Discrete covert channel}
\vspace{-\medskipamount}

As shown in section~\ref{section_NI_limits}, interferences capture information leaks in systems, but does not always mean the existence of a covert channels. What is missing in interference is the possibility to iterate a covert transmission (as in systems $S1$ and
$S2$ of Figure~\ref{interferent_examples}), but also the ability to vary the symbols (i.e. send a bit 0 or 1 ) transmitted at each interference (this is the case for system $S3$ in Figure~\ref{interferent_examples}). An immediate idea is then to define a new notion of iterated interference, with the possibility to ``transmit'' at least two distinct values at each interference, as in~\cite{Helouet03b,hzd04}. In the rest of this section, we adapt the definition of covert channels given in~\cite{hzd04} to transition systems, and show that it may still miss some obvious channels.

\begin{definition}
Let $S$ be a transition system, $q$ be a state of $S$, and $t=(q,a,q')$ be an outgoing transition of $q$. The {\em language of $S$ from $q$ after 
  $t$} is the language $L_{q,t}=\{ a.w \mid w\in \mathcal L(S_{q'}) \}$ where  $S_{q'}$ is a copy of $S$ with initial state $q'$. Let $u$ and $v$ be two users of the system. A state $q$ of $S$ is called an {\em
encoding state} for process $u$ iff there exists two distinct transitions $t_1=(q,a_1,s_1)$ and $t_2=(q,a_2,t_2)$ outgoing from $q$
and such that $Ex(a_1)=Ex(a_2) =u$, $( \Pi_v(L_{q,t_1}) \cup \epsilon) \cap \Pi_v(L_{q,t_2})= \emptyset$ and $( \Pi_v(L_{q,t_2}) \cup \epsilon) \cap \Pi_v(L_{q,t_1})= \emptyset$.
\end{definition}

More intuitively, an encoding state allows $u$ to execute two distinct actions, which consequences are disjoint and necessarily observable by $v$. Note however that discovering an encoding state is not sufficient to characterize a covert flow of information. One also needs to be able to use this kind of state (not necessarily the same at each use of the channel) an arbitrary number of times. Hence, establishing a covert channel also supposes that the sending process has means to control the system in such a way that it necessarily gets back to some encoding state. This is captured by the notion of {\em strategy}

\begin{definition}
A {\em strategy} for a user $u$ of a system $S$ is a partial function $f_u:(Q\times\Sigma\times Q)^* \longrightarrow 2^{\longrightarrow}$ that associates to every sequence of transitions $t_1.t_2\dots t_k$ of $S$ a subset of fireable transitions from the state reached after $t_k$.  
\end{definition}

In the sequel, we will only consider positional strategies, i.e. functions that only depend on the final state reached after a sequence of transitions, and we will denote by $f(q)$ the set of transitions allowed by $f$ from state $q$. A transition $t=(q,a,q')$ {\em conforms} to $f$ iff $f(q)$ is not defined or if 
$t\in f(q)$. We will denote by $S_{\mid f}$ the restriction of $S$ to transitions that conform to $f$.

\begin{definition}
\label{def_CC_discrete}
A system $S$ contains a {\em discrete channel} from $u$ to $v$ if and only if there exists two strategies $f_u$ and $f_v$ and a set of states $Q_{enc}$ such that all states $q\in Q_{enc}$ are encoding states for process $u$ in $S_{\mid f_u\cup f_v}$, and 
$f_u$ and $f_v$ allow passing infinitely often through a nonempty subset of states of $Q_{enc}$. 
\end{definition}

\begin{figure}[htbp]
\vspace{-3\medskipamount}
\begin{center}
\scalebox{0.7}{
\epsfig{figure=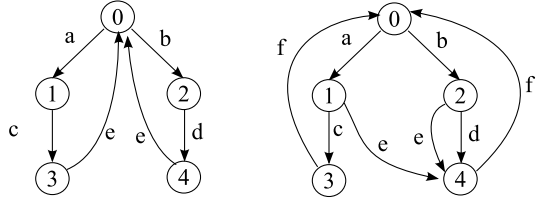, width=10cm}}
\vspace{-2\medskipamount}
\caption{Two systems with and without discrete channel}
\label{Discrete_channel_example}
\end{center}
\vspace{-2\medskipamount}
\end{figure}

Obviously, systems of Figure~\ref{interferent_examples} do not contain discrete channels. Now, let us consider the leftmost example of Figure~\ref{Discrete_channel_example}, which is composed of two users $u$ and $v$ and in which actions $a$ and $b$ are executed and observed by $u$ and all other actions are executed and observed by $v$. This system contains a discrete channel: the strategies $f_u$ and $f_v$ that allow all transitions in all states ensure that the system can pass infinitely often through the encoding state $q_0$. State $q_0$ is an encoding state as there are two transitions  $t_1=(q_0,a,q_1)$ and $t_2=(q_0,a,q_2)$ fireable from $q_0$, and such that $(\Pi_v(L_{q_0,t_1}) \cup \epsilon) \cap
\Pi_v(L_{q_0,t_2})=\emptyset$ and  $(\Pi_v(L_{q_0,t_2}) \cup \epsilon) \cap
\Pi_v(L_{q_0,t_1})=\emptyset$. Hence, the consequences of firing $t_1$ or $t_2$ that can be observed by $v$ do not contain the empty word, and are disjoint. 
It then suffices for instance to perform action $a$ from state $0$ to pass a bit $0$ and to perform action $b$ to pass bit $1$ from $u$ to $v$. Note that our definition of discrete channel makes no supposition on the code established by the two corrupted users, but only ensures that at least two distinct choices of $v$ have distinct observable consequences for $v$. 

\begin{figure}[htbp]
\vspace{-2\medskipamount}
\begin{center}
\scalebox{0.7}{
\epsfig{figure=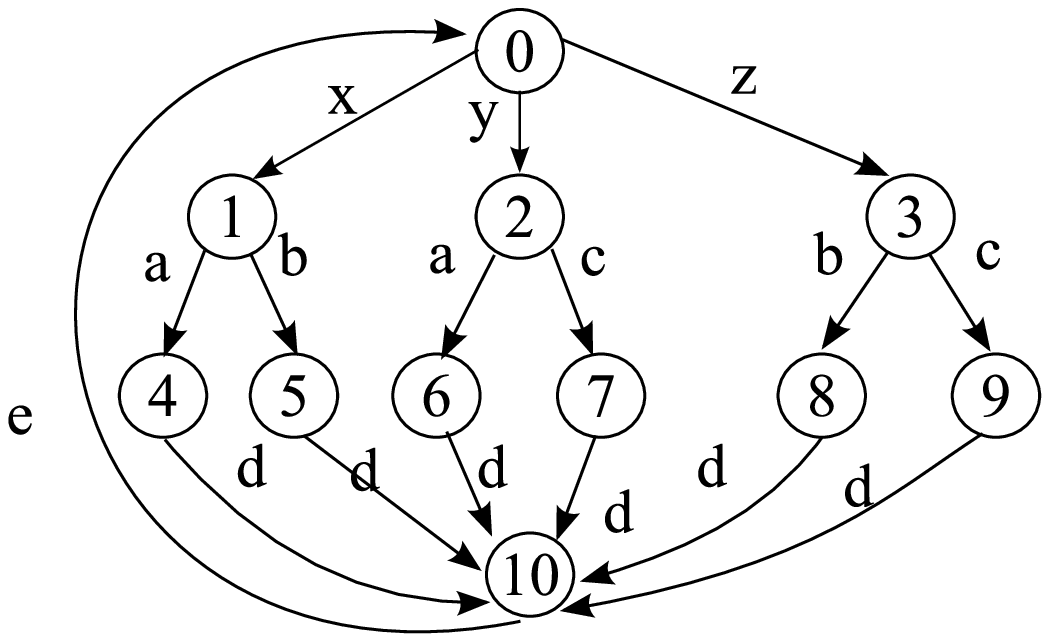, width=7cm}\hspace{1cm}\epsfig{figure=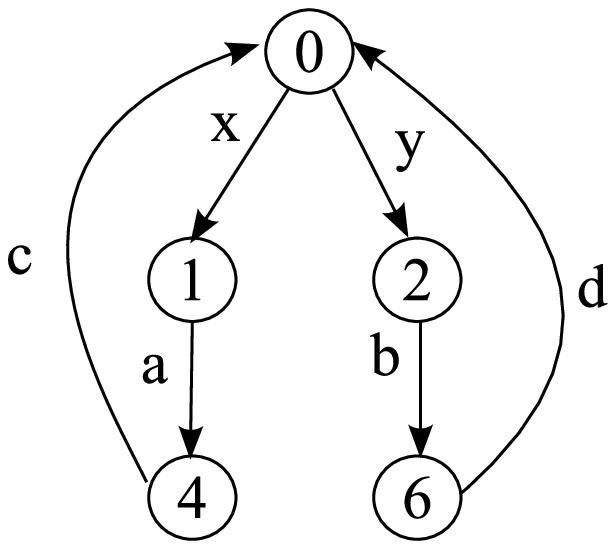, width=4.5cm}}\hspace{1cm}\scalebox{0.7}{\epsfig{figure=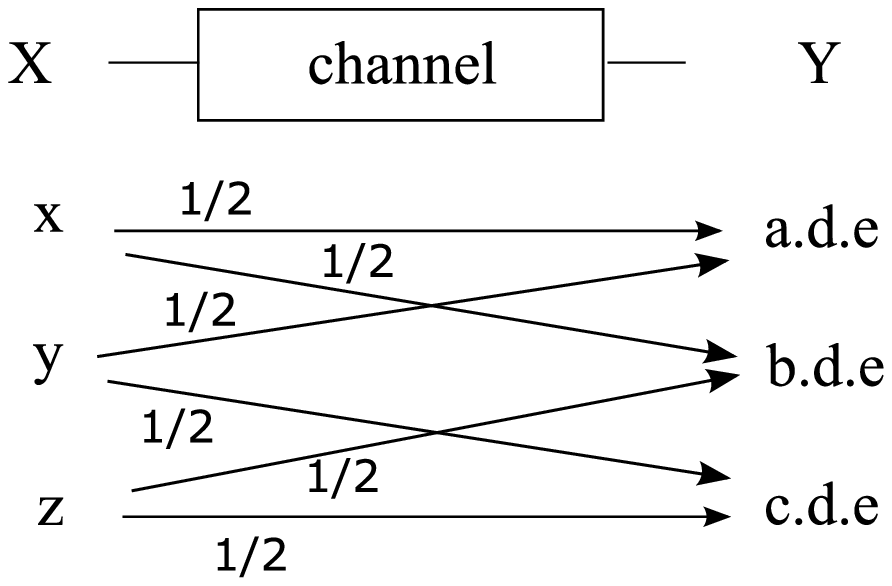,width=6cm}}
\vspace{-\medskipamount}
\caption{a) A covert channel free system? b) No covert channel c) The equivalent channel of Fig.~\ref{CC_discrete}-a}
\label{CC_discrete}
\end{center}
\vspace{-3\medskipamount}
\end{figure}

Let us consider the rightmost system of Figure~\ref{Discrete_channel_example}, comporting three agents $u,v$ and $r$, and such that actions  $a$ and $b$
are executed and observed by $u$, actions $c,d$ and $e$  executed by $r$ and observed by $r$ and $v$, and action $f$ is executed and observed by $v$. This system does not contain a discrete channel from $u$ to $v$, as $\Pi_v(L_{q_0,t_1}) \cap \Pi_v(L_{q_0,t_1}) = e.f((c+e+d).f)^*$.
If we consider definition~\ref{def_CC_discrete}, we can notice that the kind of covert channel considered by the definition uses the control flow of a system to transfer information. % Covert channel are frequently classified as ``storage channel'' when they use a resource of the system to store and then read bits of a covert message, or as ``timing channels'' when time is used as covert information vector. The control flow channels of definition~\ref{def_CC_discrete} would fall into the class of storage channels, even if the global state of a system can not really be considered as a resource.
However, this approach only detects control flow channels in which more than one bit of information is transferred at each use of the channel. Efficient communication techniques allow communications over channels event when a fragment of bit is sent at each use of the channel, and covert channel can clearly benefit from efficient coding and decoding schemes. Consider for instance the example of Figure~\ref{CC_discrete}-a. Actions  $x,y$ and $z$
are executed and observed by an agent  $u$, actions $a,b$ and $c$ are executed by an agent $r$ and observed by $r$ and $v$, and last actions $d$ and $e$ are executed and observed by agent $v$. According to definition~\ref{def_CC_discrete}, this system contains no discrete channel from $u$ to $v$, because for any choice of a transition $t_i$ by $u$ from state $q_0$, there exists another choice $t_j$ such that the observable consequences of $t_i$ on agent $v$ are not disjoint from the consequences of $t_j$. Now, let us consider more precisely the sequences of actions that can be seen by agent $v$. When $v$ observes a word $a.d.e$, he can not know whether the former choice of $u$ was $x$ or $y$, but he knows for sure that this choice was not $z$. This situation is similar for every word observed by $v$ between two choices of $u$. Hence, at each use of this system, agent $u$ can send ``not x'', ``not y'', or ``not z'' that is three distinct values to agent $v$, even if the system does not contain discrete channels in the sense of definition~\ref{def_CC_discrete}.

The system of Figure~\ref{CC_discrete}-a can be seen as a simple communication channel, in which user $u$ sends symbols from the input alphabet $\mathcal X=\{x,y,z\}$ and user $v$ receives symbols from the output alphabet $\mathcal Y=\{a.d.e, b.d.e, c.d.e\}$. This communication channel is represented in Figure~\ref{CC_discrete}-c, and its capacity is $0.5849$  bit per use of the covert flow (details are omitted here, but can be found in the extended version). 
%
%
%
% $$\begin{array}{clcccccc}
% C&=\underset{\mathcal P(X)}\max &H(X) &+ H(Y)& -H(X,Y)\\
%
% &= \underset{\mathcal P(X)}\max& -\underset{x\in\mathcal X}\Sigma p(x).log_2(p(x)) &- \underset{y\in\mathcal
%   Y}\Sigma p(y).log_2(p(y))&+ \underset{x,y\in\mathcal X\times\mathcal Y}\Sigma
% p(x,y).log_2(p(x,y))\\
%
% &=  &-3.\frac{1}{3}/\log_2(\frac{1}{3}) & -3.\frac{1}{3}/\log_2(\frac{1}{3})
% &+ 6.\frac{1}{6}/\log_2(\frac{1}{6}) \\
%
% &=0.5849
% \end{array}$$
%
% \begin{figure}[htbp]
% \begin{center}
% \epsfig{figure=CC_discrete_TI.eps,width=6cm} 
% \end{center}
% \vspace{-2\medskipamount}
% \caption{The commuication channel hidden in the system of Figure~\ref{CC_discrete}-a} 
% \label{CC_discrete_TI}
% \vspace{-3\medskipamount}
% \end{figure}
%
The translation from the transition system Figure~\ref{CC_discrete}-a to the channel Figure~\ref{CC_discrete}-c is straightforward, and the non null capacity shows the presence of a covert flow in the system. 
%Remark that each use of the channel passes less than one bit of information from $u$ to $v$. This is why the characterization of definition~\ref{def_CC_discrete} does not detect this flow. 
Remark that this channel passes less than one bit of information from $u$ to $v$ at each use, and hence is not captured by definition~\ref{def_CC_discrete}.
%In section~\ref{section_characterization},  we will characterize covert flows 
%with capacities $<1$ for a certain kind of transition systems. 

\vspace{-2\medskipamount}
\subsection{First use of information theory to discover information leaks}
\vspace{-\medskipamount}

This paper is not the first attempt to use information theory to discover information leaks. Millen~\cite{Millen87} considers a machine model that accepts inputs from agents and produces outputs. Inputs are chosen by several users from an alphabet $I$, and after a sequence of inputs $w\in I^*$, every agent $u$ in the system can observe some outputs, denoted $Y_u(w)$. If we denote by 
$X_u(w)$ the sequence of inputs performed by agent $u$ during input sequence $w$, and by $\bar\pi_{X_u}(w)$ the projection of $w$ on inputs performed by agents other than $u$, then a non-interference property between $u$ and $v$ can be written as $\forall w\in I^*, Y_v(w)=Y_v(\bar\pi_{X_u}(w))$.
Millen then shows that if $u$ and $v$ do not interfere, then if all inputs of users other than $u$ are independent from $X_u$, the mutual information $I(X_u;Y_v)$ between $X_u$ and $Y_v$ should be null. Note that this is only an implication, and that there might be cases where $I(X_u;Y_v)=0$, but nevertheless  $u$ interferes with $v$. The converse property is more interesting: a non null mutual information between $X_u$ and $Y_v$ denotes an interference. The mutual information $I(X_u;Y_v)$ is computed over input sequences performed by $u$ between two outputs to $v$. 

The machine model is not explicitly given in this approach: it is a black box that receives a {\em finite} sequence of inputs from all users and produces a {\em finite} sequence of outputs. Within the context of transition systems, the inputs $X_u$ of agent $u$ are the actions that $u$ can execute ($Ex^{-1}(u)$), and the outputs to $v$ the actions that $v$ can observe. Note however that the independence hypothesis between inputs of the system does not necessarily hold, and furthermore that nothing guarantees that an output to $v$ happens after a {\em finite number} of inputs from $u$, nor that outputs are finite. Hence, computing $I(X_u;Y_v)$ means computing the mutual information between sets of inputs/outputs of arbitrary size. Without giving any hint on how to compute this value, Millen's characterization of information leaks can be rewritten as follows for transition systems. The average information leak (per transition) between two agents $u$ and $v$ in system $S$ is $Leak(u,v)= \underset{n\rightarrow \infty}\lim~~ \frac{1}{n}.~\underset{ P(\mathcal L^n(S))}\max ~~ I\left(\Pi_{Ex^{-1}(u)}(\mathcal L^n(S));\Pi_v(\mathcal L^n(S))\right)$, where $\mathcal L^n(S)$ is the set of words in $\mathcal L(S)$ of length $n$. 
If $Leak(u,v)>0$, then there is an interference from $u$ to $v$.
However, this average mutual information defines a correlation between actions of $u$ and observations of $v$, which can be called average interference, but {\em is not} a characterization of a covert channel. Indeed, we can show on a very simple example that non-zero value for $Leak(u,v)$ implies an average interference greater than zero, but not a deliberate information flow between two users. Consider for instance the system of Figure~\ref{CC_discrete}-b. Let $a,b$ be the actions executed and observed by $u$, $x,y$ the actions executed and observed by $r$ and $c,d$
the actions executed and observed by $v$. In this example, successive actions and observations of processes $r,u$ and $v$ are 
independent. If we suppose that choices of process $r$ are equiprobable, then the average information leak is $1/3$ bits at each transition (the details of the calculus are provided in the extended version). 
%
%
% $u$ and $v$ have no possible maximization on any probabilistion distribution, and $H\left(\Pi_{Ex^{-1}(u)}(\mathcal L^n(S))\mid \Pi_v(\mathcal
%   L^n(S))\right)=0$.  We then have:
%
% $$\begin{array}{llll}
% &\underset{n\rightarrow \infty}\lim~~ \frac{1}{n}. \underset{ P(\mathcal L^n(S))} \max I\left(\Pi_{Ex^{-1}(u)}(\mathcal L^n(S));\Pi_v(\mathcal
%   L^n(S))\right)&
% =&\underset{n\rightarrow \infty}\lim~~ \frac{1}{n}. H\left(\Pi_{Ex^{-1}(u)}(\mathcal L^n(S)) \right) \\
% = &\frac{1}{3} . H\left(\Pi_{Ex^{-1}(u)}(\mathcal L^3(S))\right)&
% = & \frac{1}{3}.\left ( - \frac{1}{2}\log_2(\frac{1}{2}) - \frac{1}{2}\log_2(\frac{1}{2})\right) = \frac{1}{3}
% \end{array}$$
%
This means that user $v$ infers an average number of $1/3$ bits on $u$'s behavior at each transition of the system. There is clearly an interference in the system, as user $v$ can almost infer the exact behaviors of users $u$ (up to the last $a$ or $b$) from his own observations. Furthermore, such kind of interference can occur an arbitrary number of times. 
%
%INSERT
% 
Let us recall at this point a major difference between covert channels between two agents $u,v$ and interference between $u$ and $v$. When an interference occurs, process $v$ has learned some information about $u$'s actions or states. In a covert channel, $u$ {\bf decides} to send some hidden information to $v$, and both processes have agreed on a protocol to convey this information. Hence, we can not consider that the system of Figure~\ref{CC_discrete}-b contains a covert channel from $u$ to $v$ (nor the converse direction), as the decision to execute $a$ or $b$ is not a choice from process $u$, but a consequence of a choice of process $r$ (that chooses between action $x$ and $y$). However, a covert channel exists from $r$ to $u$ and from $r$ to $v$, as it suffices for process $r$ to play action $x$ or $y$ to allow distinct observations on $u$ and on $v$.
The covert channel 
characterization introduced in section~\ref{section_characterization} shows a dependency between the {\em decisions of a sending process}, that is actions performed from states in which this process has more than one fireable action, and {\em observable consequences on a receiving process}. However, even reformulated this way, nothing guarantees that the capacity of such covert channels can be expressed as a closed formula.

% Taking this into account, the capacity of (\ref{Millen_equation_adapted}) for information flows rewrites as:

% \beq C_F(u,v)&=\underset{n\rightarrow \infty} \lim \frac{1}{n} . \underset{ P(In^n)} \max I \left(In^n;Out^n)\right)>0,\mbox{~where~}\label{flow_Capacity}\eeq

% $$\begin{array}{lll}

% In^n=\{ \Pi_{Ex^{-1}(u)}(w)\mid& w=v.a\in \mathcal L^n(S) \wedge \exists q,q',q''\in Q, a,b\in Ex^{-1}(u), \\
% &q_0\overset{v}\Longrightarrow q, q\overset{a}\longrightarrow q',q\overset{b}\longrightarrow q''  \}\\
% \\
% Out^n=\{ \Pi_{Obs^{-1}(v)}(w)\mid& w=u.a.v\in \mathcal L^n(S) \wedge \exists q,q',q''\in Q, a,b\in Ex^{-1}(u), \\
% &q_0\overset{u}\Longrightarrow q, q\overset{a}\longrightarrow q',q\overset{b}\longrightarrow q''  \}
% \end{array}$$

% Intuitively, $In^n$ lists all sequences of actions of $v$ that are executed along a path of length $n$ that contains at least one decision of $u$  (i.e it contains all choices of $u$). Similarly, $Out^n$ lists all possible observed actions on process $v$ along a path of length $n$ containing a choice of process $u$. Note however that if the definition of such capacity is rather straightforward, computing it is not an easy problem, as it involves finding a limit, that is computing a maximal mutual information over all probabilizations of $In^n$. Furthermore, alphabets $In^n$ and  $Out^n$ grow rapidly with $n$, and the maximization problem rapidly becomes intractable. We will show in the sequel situations where capacity have closed form, or is defined as an expression that converges rapidly.

Several other works have used information theory to compute the bandwidth of already identified covert flows. I.~Moskowitz shows in~\cite{Moskowitz94} how a very simple information aggregation system can be perverted to create a covert channel between two users, and computes its capacity. We will adapt this example in section~\ref{section_example} to illustrate our characterization of covert channels. Several other works have studied and quantified covert flows in TCP/IP~\cite{Ahsan02,MurdochL05}. Some channels simply consist in filling useless bits in TCP frames to hide information (this security leak called TCP piggybacking has been corrected ever since). More elaborate coding strategies modulate the use of packets in time windows to pass information (the shape of the function denoting the cumulated number of messages encodes the input symbols in a covert channel)~\cite{Giles02}. I. Moskowitz remarks that adding non-deterministic delays for the transmission of packets reduces the capacity of timing covert channels, and has proposed a mechanism called the ``network pump''. The pump stores sent packets for a random time, and then delivers them to their destination~\cite{Kang96}. This mechanisms does not close covert flows, but reduces their capacity, as demonstrated by~\cite{Giles02}. However, it also impose a time penalty to honest users of the system.

\begin{figure}[htbp]
\vspace{-2\medskipamount}
\begin{center}
\scalebox{0.7}{
\epsfig{figure=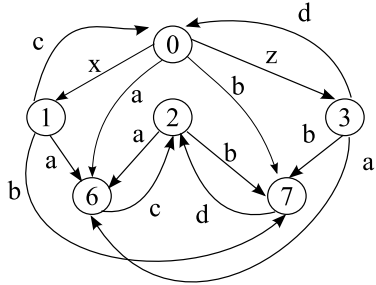, width=7cm}}
\vspace{-\medskipamount}
\caption{A system that is not interferent, but contains a covert channel}
\label{CC_but_no_Interf}
\end{center}
\vspace{-3\medskipamount}
\end{figure}

%INSERT *5
At this point, let us summarize the mentionned differences between interference and presence of covert flows in systems. Systems that contain a leak but do not allow iteration of leaks, or that leak the same information from an agent to another an arbitrary number of times do not contain covert flows, but are interferent. A generalization of interference to iterated multivalued interference (definition~\ref{def_CC_discrete} in this paper) fails to characterize covert flows of information with capacity lowr than $1$. When interference is characterized as a non-null mutual information between actions of an agent and observations of another (as in \cite{Millen87}), some systems might be found interferent, bu yet do not contain covert flows. Last, consider the example of Figure~\ref{CC_but_no_Interf}, where actions $a,b$ are executed and observed by agent $u$, actions c,d executed and observed by agent $v$, and actions $x,y$ are executed and observed by agent $r$. In this system, $u$ does not interfere with $v$ (both if we consider SNNI or BSNNI), but this system contains a cover channel, as all $a$'s are followed by a $c$, and all $b$'s are followed by a $d$. We do not know if such situation ecapes a characterization by all notions of interference, but so far, it seems that although interference and covert flow presence look very similar, they are indeed {\em orthogonal properties} of systems. 
 
\vspace{-2\medskipamount}
\section{An example: the acknowledgment channel}
\label{section_example}
\vspace{-\medskipamount}

We have shown in previous sections that a characterization of covert channels should 1) highlight dependencies between iterated and deliberate choices of one sending process on one side, and the observed consequences of these choices on another process, but also that 2) each occurrence of these choices can allow the transfer of less than one bit of information. Language theory is well suited to deal with $1)$ while information theory is well adapted to deal with $2)$. Section~\ref{section_characterization} reconciles both parts of the problem, and brings covert channel discovery back to the computation of a capacity for channels with side state information. 

\begin{figure}[htbp]
\vspace{-2\medskipamount}
\begin{center}
\epsfig{figure=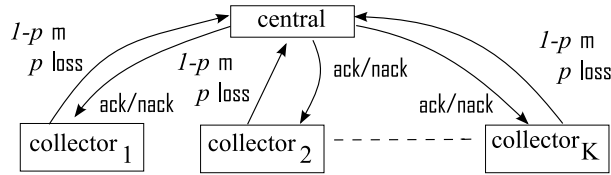, width=8cm}
\vspace{-2\medskipamount}
\caption{An information aggregation system.}
\label{example_moskowitz}
\end{center}
\vspace{-4\medskipamount}
\end{figure}

Let us illustrate our approach with a toy example inspired from~\cite{Moskowitz94}. Figure~\ref{example_moskowitz} describes an information aggregation system. Several agents called {\em collectors} collect information that are then sent to another agent called the {\em central observer}. The role of the central observer is to analyze the data provided by observers, and to give an overall expertise (statistics, computation of mean values, ...). Data are sent at regular rate from each collector to the central, using data messages of type $m$. The ascending communication path is not reliable, and can lose messages, with probability $p$. The descending path is reliable, and the central sends acknowledgment messages to its collectors to indicate whether it received or not the last message. Message losses are detected using timeouts. When a message is successfully received, the central sends an $Ack$ message to the collector. In case a timer expires, the central sends a $Nack$ message with the expected packet number to the collector. Collector wait for the answer from the central before sending a new packet, that is at a given time, there is at most one data packet transiting from each collector to the central. 

In addition to this simple mechanism, the system has strict security rules: collectors have no means to communicate, and the central can not communicate with collectors. The main reason is that the individual data collected locally has almost no value, but that the global data obtained by aggregation or the diagnosis computed should be protected. For this reason, communications from the central to the collectors are observed, and a monitor filters all messages that are not $Ack/Nacks$. In addition to this, the central observer is frequently audited, and should have received all acknowledged packets. An immediate idea to establish a covert channel is to use $Ack$ and $Nack$ messages to transfer $0$ or $1$ from the central to a chosen collector. However, the central can not declare as received a lost packet, as this will be discovered during the audit. Nevertheless, when a packet is received, it can be falsely declared as lost, and a useless retransmission is done by the collector.

\begin{figure}[htbp]
\begin{center}
\epsfig{figure=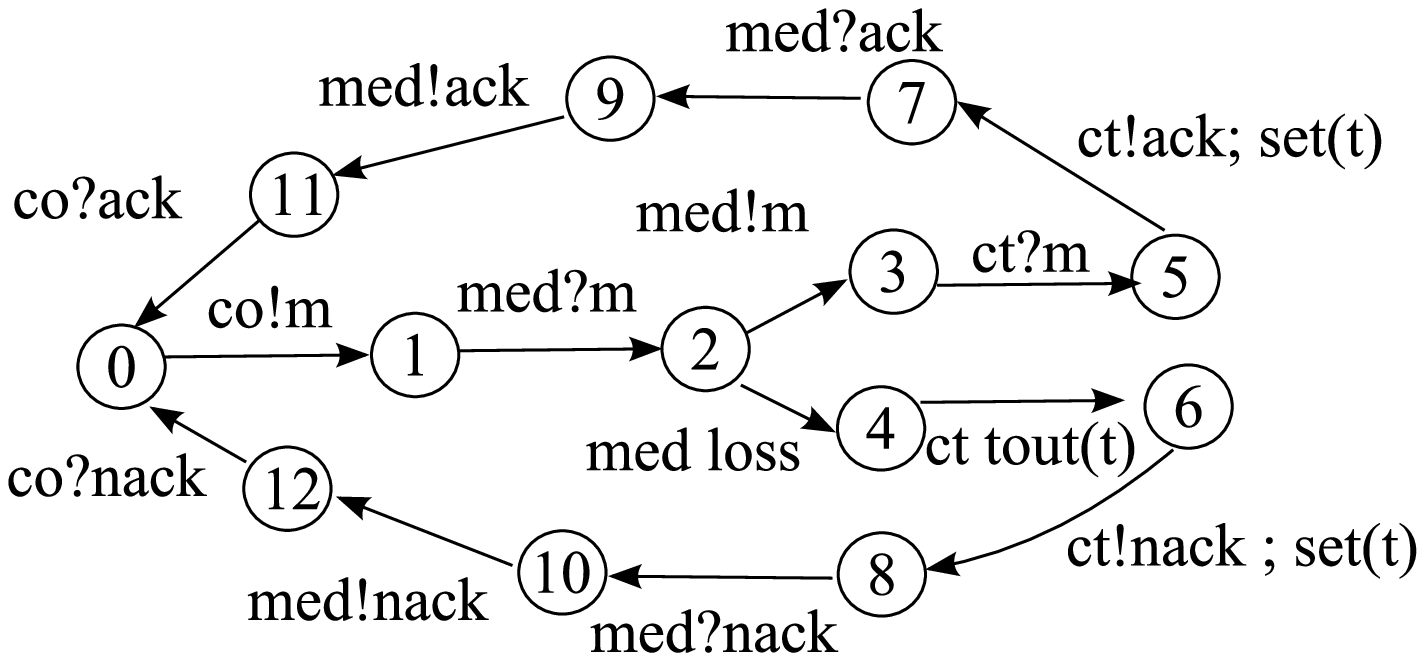,width=7cm}\hspace{0.5cm}\epsfig{figure=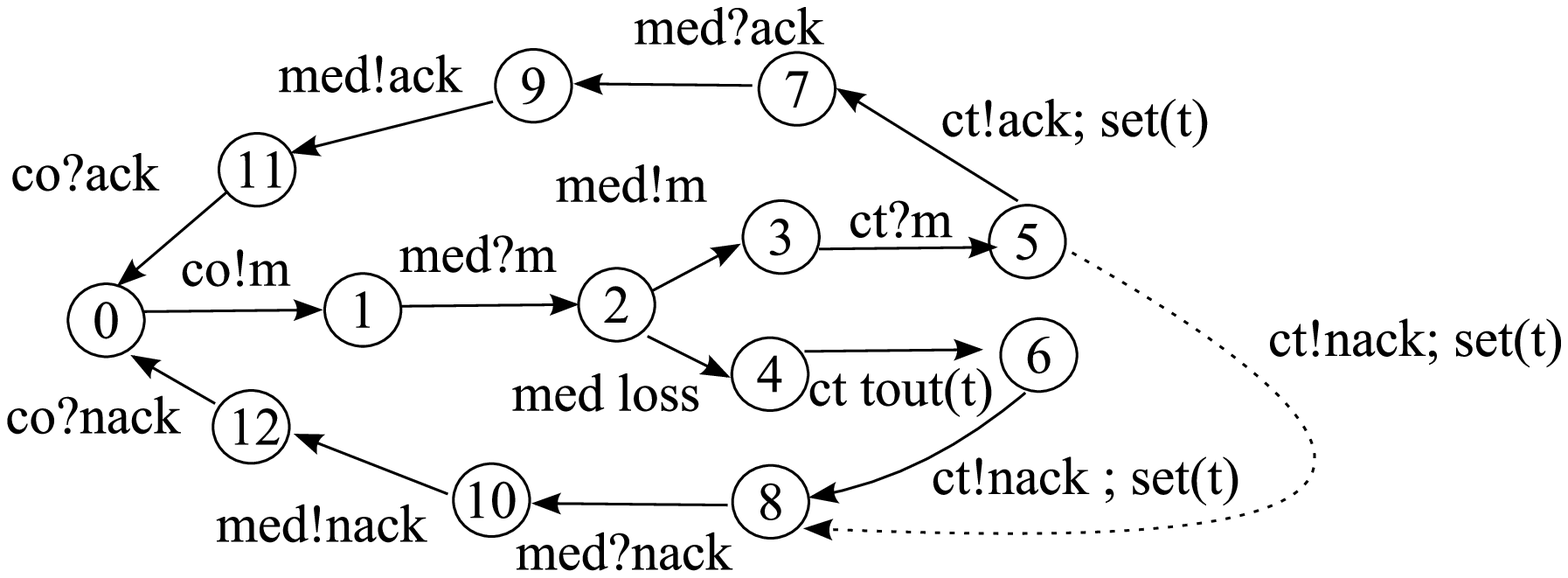,width=9cm}
\vspace{-3\medskipamount}
\caption{a) The original protocol b) The corrupted protocol to build a covert channel}
\label{ex_mosk_system}
\end{center}
\vspace{-3\medskipamount}
\end{figure}

Figure~\ref{ex_mosk_system}-a depicts the normal behavior of a pair collector/central, i.e. when none of them have been corrupted to establish a covert information flow. The lossy communication medium has been represented as an additional process, that simply transfers $Ack$ and $Nack$ messages, and transfers or looses $m$ messages along the ascending way. To depict this system, we will consider three processes: $ct$, the central observer, $med$, the communication medium, and $co$ the collector. Actions of the system are of the form $p!x$, $p?x$, $p~loss$, $p~set$ and $p~tout$, denoting respectively the sending of a message of type $x$ by a process $p$, the reception of a message of type $x$ by a process $p$, the loss of a message, a timer setting or a timeout. In addition to the transitions, we add to the model a function $P_S$ that associates probability $p$ to the loss of message $m$. We hence have $P_S(2,med!m,3)=(1-p)$ and $P_S(2,med loss,4)=p$. This system can not be used to transfer information from $CT$ to $CO$, as $CT$ just copies the choices of the communication medium.

Figure~\ref{ex_mosk_system}-b shows a slightly modified system, where the central observer can cheat, and declare lost a packet it has received. With this new protocol, when the communication medium does not loose the transmitted packet (this occurs with probability $1-p$), the central observer can pass without loss or noise a bit of information (sending $ack$/$nack$ eventually leads to a reception of the acknowledgment message) to the collector. This is a perfect channel. Conversely, when a packet is lost (this occurs with probability $p$, the central observer can only send a $nack$ packet, that will eventually be received by the collector. This is a zero capacity channel. 
%If we consider this system from the covert channel point of view, we can notice that the system is a random alternation of two communication channels. 
If we look at this system from the covert channel point of view, it is a state channel, where the state only depends on the medium (and not on the inputs/outputs ($ack$/$nack$)) and is iid. Moreover, the central knows causally in which state the system is. 
We can then apply the technique of~\cite{Shannon58} described in Thm 1 to compute the capacity $C_{Mosk}(CT,CO)$ of this channel.

% \begin{figure}[htbp]
% \begin{center}
% \epsfig{figure=ex_mosk_system_perverted.eps,width=10cm}
% \caption{The corrupted protocol to build a covert channel}
% \label{ex_mosk_system_perverted}
% \end{center}
% \end{figure}

% In this situation, the central knows exactly in which sate the system is. We then have a two state channel with side information at the transmitter, that can be represented with the model of Figure~\ref{ex_mosk_channel}. 
% One can easily notice that in this example, according to the state in which the system is located, $CT$ and $CO$ communicate either through a perfect or zero-capacity channel. In these channels, the input symbols used by $CT$ are of course the sending of ack/nack messages, and the output symbols read by $CO$ are the receptions of ack/nack.

\begin{figure}[htbp]
\begin{center}
\begin{minipage}[l]{0.46\linewidth}
\scalebox{0.7}{
\epsfig{figure=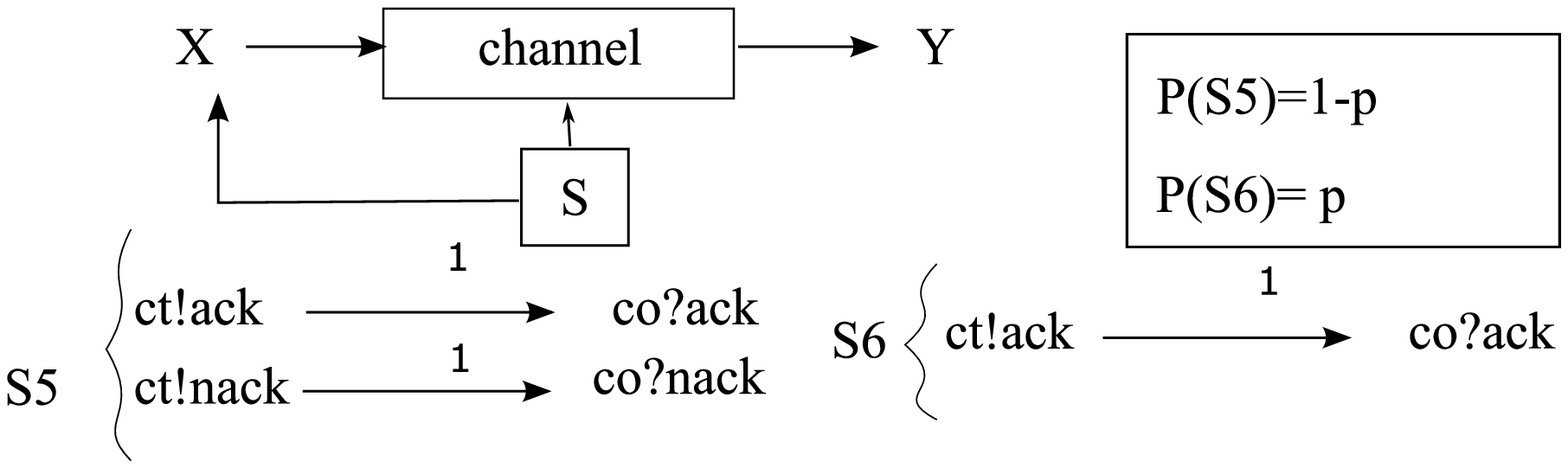,width=10cm}\hspace{1.5cm}\epsfig{figure=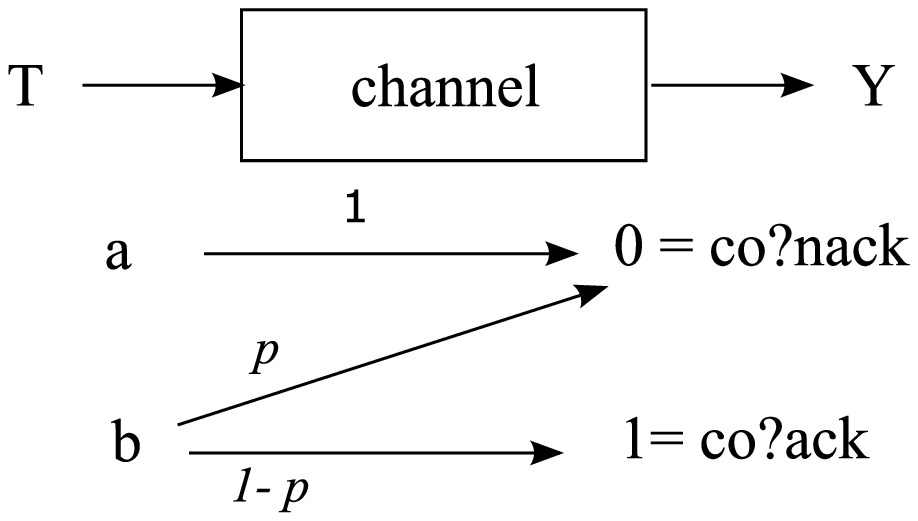,width=5cm}}
\end{minipage}
\begin{minipage}[r]{0.46\linewidth}
\scalebox{0.8}{\hspace{5cm}
$\begin{array}{c|cc}
    T & P(Y=0\mid T) & P(Y=1\mid T)\\
    \hline
    a & 1 & 0\\
    b & p & 1-p
    \end{array}$}
\end{minipage} 
\vspace{-\medskipamount}
\caption{a) A state channel for the flow from $CT$ to $CO$ b) A stateless channel with identical capacity}
\label{ex_mosk_channel}
\end{center}
\vspace{-3\medskipamount}

\end{figure}

Figure~\ref{ex_mosk_channel}-a shows our two-state channel with side information. According to the state in which the system is, $CT$ and $CO$ communicate either through a perfect or zero-capacity channel. When in state $S5$, the input symbols used by $CT$ are the sending of $ack$/$nack$ messages, i.e. $X_{S=S5} \in \{!ack,!nack)\}$  and when in state $S6$, $X_{S=S6} \in \{!ack\}$. Whatever the state is, the output symbols read by $CO$ are the receptions of $ack$/$nack$, $Y\in\{1=?ack,0=?nack)$.
From Thm 1, the state channel (see Figure~\ref{ex_mosk_channel}-a) is equivalent to a stateless channel depicted in Figure~\ref{ex_mosk_channel}-b. Its input is a 2-length vector $T=(X_{S=S5},X_{S=S6})$ that can take 2 values:  $T \in \{ a=(!nack,!nack), b=(!ack,!nack)\}$ and the output alphabet is unchanged. The transition probabilities (see table in Figure~\ref{ex_mosk_channel}-b) are those of the so-called Z-channel~\cite{CoverThomas}. Therefore the capacity of the equivalent stateless channel is  $C_{Mosk}(CT,CO)=\log_2(1+2^{-s(p)})$, where $s(p)=\frac{-p \log p -(1-p)\log (1-p)}{1-p}$. It is interesting to compare this capacity $C_{Mosk}(CT,CO)$ with the case with perfect state knowledge at the collector as well. With perfect state knowledge at both central and collector, the capacity is $1-p$ since 1 bit of information is sent each time the channel is in state $S5$, which occurs with probability $1-p$. In our example, only the transmitter knows the state and the capacity $C_{Mosk}(CT,CO)\le 1-p$.

\vspace{-2\medskipamount}
\section{IT based characterization of covert flows}
\label{section_characterization}
\vspace{-\medskipamount}
%
% In this section, we generalize the case study of section~\ref{section_example}. The main idea is to establish a correspondence between a subclass of transition systems called {\em Half-duplex}, and communication state channels defined in theorem~\ref{thm_state_channel_capacity}. Then, computing a covert flow capacity resumes to computing the capacity of a communication state channel.  
%
%INSERT *6
% 
We have shown in previous sections that the terms interference and covert channels refer to different kinds of leaks, and orthogonal properties of systems. Similarly, discrete channels characterization misses some obvious channels with capacity lower than $1$. This section proposes an information theoretic characterization of covert flows. The main idea is to consider that a hidden channel exists from user $u$ to user $v$ in a system $S$ if $u$ and $v$ can use $S$ to simulate a memoryless communication channel with state with capacity greater than $0$. We define this characterization in two steps: we first define some transition systems that simulate a communication channel with state. These systems will be called {\em Half-Duplex} systems. We then define control flows covert channels as the capacity for a pair of users $u,v$ to change their interactions with the rest of the system, i.e. transform a system $S$ into a new system $S'$ in such a way that $S'$ is a Half-duplex system, and is observationnaly equivalent to $S$ for all other agents. Then, computing the capacity of such covert flow resumes to computing the capacity of the communication channel simulated by $S'$.

\vspace{-\medskipamount}
\begin{definition}
Let $S=(Q,\longrightarrow,\Sigma,q_0)$ be a transition system. A state 
$q\in Q$ is  {\em controlled} by process  $p\in \mathcal P$ iff for any transition $(q,a,q')$, $Ex(a)=p$. The system $S$ is in {\em Half-Duplex} between two processes  $u$ and $v$ iff there exists a state $x$ (called the {\em control state}), a set of states $EN = q_1,\dots, q_n$ controlled by $u$ (called the {\em encoding states}) and two bounds $K_1,K_2$ such that:
\vspace{-0.5\medskipamount}
\begin{itemize}
\item any path originating from $x$ of length greater or equal to $K_1$ passes
through one state of $EN$,
\vspace{-\medskipamount}
\item  any path originating from a state  $q_i\in EN$ of length greater or equal to $K_2$ passes through $x$ and contains at least one transition labeled by an action in $Obs^{-1}(v)$,
\vspace{-\medskipamount}
\item no path from $x$ to one of the states in $EN$ uses transitions labeled by $Ex^{-1}(u)$.
\end{itemize}
\vspace{-0.5\medskipamount}

For simplicity, we will also assume that $Y=\Pi_{Obs^{-1}(v)}\left( \mathcal L(Path(x,x))\right)$ forms a code, i.e. any word $w$ in $Y^*$ has a unique factorization $w=Y_1.Y_2\dots Y_k$.

\end{definition}

More intuitively, the states of $EN$ are states from which the sending process $u$ in a covert channel will encode information. Passing from state $x$ to a state $q_i\in EN$ simulates a choice of a new channel state, and is an essential condition to allow for a translation of transition systems into state channels. This does not mean however that the capacity of hidden flows can not be computed in transition systems that do not meet this condition. The additional condition that all states in $EN$ are controlled by $u$ is not a real constraint. In fact, we can easily ensure that all states are controlled by a single process by adding an additional actor to the system that schedules the next process allowed to move.
Clearly, Half-duplex channels simulate the behavior of communication channels with side state information, and the states of this channel will be the encoding states of the half-duplex system. 
% INSERT
Chosing an action (or a sequence of actions) of $u$ from an encoding state $q_i$ simulates the sending of an input symbol in the communication channel with state, and the observation performed by $v$ before returning to state $x$ simulates the reception of an output symbol. 
With the constraints imposed by the definition, we can ensure that the choice of an encoding state is an i.i.d variable, that is independent from actions of $u$ and $v$. The code assumption means that the receiving process in the covert flow also knows how many times the channel was used. This hypothesis can be easily relaxed, but we will use it in the rest of the paper as it simplifies translation to a communication channel model. We refer interested 
readers to the extended version of this work for a more general framework without this code assumption. 

\begin{figure}[htbp]
\begin{center}
\epsfig{figure=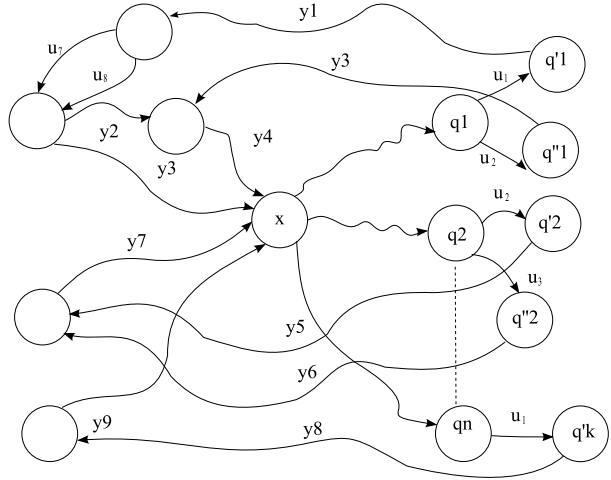, width=8cm}
\vspace{-2\medskipamount}
\caption{General shape of a Half-duplex transition system}
\label{half_duplex_figure}
\end{center}
\vspace{-2\medskipamount}
\end{figure}

Figure~\ref{half_duplex_figure} shows the general shape of half-duplex systems. There is a single concentrator state $x$. Unlabelled arrows originating from $x$ denote a sequence of transitions that is unobservable for $v$. Arrows labeled by $u_i$ denote an action of process $u$. Arrows labeled by $y_i$ denote sequences of transitions for which word $y_i$ is observed by process $v$.
Half-duplex transition systems simulate the behavior of some communication channel with state. We can easily build the simulated channel, and then compute the capacity of the obtained model to quantify the importance of a leak.  

\begin{definition}
Let $S=(Q,\longrightarrow,\Sigma,q_0)$ be a Half-duplex transition system from $u$ to $v$, with encoding states $EN$, control state $x$, and probability function $P_S$. The {\em state channel from $u$ to $v$} contained in $S$ is the state channel with side information $Ch_{S,u,v}=(S,X,Y,g_{s},\{p_s(y\mid x)\}_{s\in \mathcal S})$, where:
\vspace{-0.2\medskipamount}
\begin{itemize}
\item $S$ is a random variable over a set of states $\mathcal S =EN$
\vspace{-\medskipamount}
\item $g_{q_i} = \Sigma_{\rho \in Path(x,q_i)} P_S(\rho)$ is the 
probability to reach state $q_i$ from state $x$. 
\vspace{-\medskipamount}
\item $X$ is a random variable defined over $\mathcal X=\underset{q_i\in EN}\bigcup \Pi_{Ex^{-1}(u)}\left(\mathcal L(Path(q_i,x))\right)$, the set of sequences of actions executed by process $u$ to move from a state $q_i \in EN$ to state $x$. 
\vspace{-\medskipamount}
 \item $Y$ is a random variable defined over $\mathcal Y = \underset{q_i\in EN}\bigcup \Pi_{Obs^{-1}(v)}\left (\mathcal
  L(Path(q_i,x))\right)$, the set of sequences of actions observed by process  $v$ to go from state $q_i$ to state $x$. 
\vspace{-\medskipamount}
\item $p_{q_i}(y\mid w)=\underset{\rho \in Path(q_i,x), 
\Pi_u(\mathcal L(\rho)) = w,\Pi_v(\mathcal L(\rho)) = y} \sum P_S(\rho)$
\end{itemize}
\vspace{-\medskipamount}

\end{definition}

The translation from a half duplex transition system to a state channel immediately gives a capacity of information flows from $u$ to $v$. Note however that this implicitly means that the sending process in the covert channel must have perfect knowledge of the system's state to achieve this capacity. 
When process $u$ does not know perfectly the state of the system, the capacity of the state channel computed from $S$ should only be seen as an upper bound of an achievable capacity.

The capacity $C_{S,u,v}$ of the state channel computed from a half duplex transition system $S$ gives the average mutual information between sequences of choices executed by $u$ and observations of process $v$ between two occurrences of state $x$. Non-zero capacity of such information flow from $u$ to $v$ in a Half duplex system then highlights a capacity to transfer information from $u$ to $v$ using system $S$. Note however, that a capacity is an average number of bits {\em per use} of the channel, and does not give the bandwidth of the channel, which is an average number of bits transferred {\em per time unit}. Usually, capacity and bandwidth are tightly connected for communication channels, as it is frequently assumed that channels are used at a constant rate $T$. However, in our covert flow setting, each channel use may have a distinct duration. We do not yet know whether there exists a closed form or single letter characterization for the bandwidth of a covert channel. However, non-zero capacity means non-zero bandwidth, and
capacity is still relevant to characterize covert flows.

Now that Half-duplex systems are defined, we can propose a characterization for some covert channels. The main intuition behind the following definition is that a covert flow exists when two users $u,v$ can modify their behavior in such a way that the resulting system is observationnally equivalent for all other agents, but simulates a communication channel with state from $u$ to $v$, with capacity greater than $0$.

\begin{definition}
\label{def_CC}
A transition system $S$ contains a {\em control-flow covert channel} from $u$ to $v$ if communications from $u$ to $v$ are not allowed by the security policy, and there exists a Half-duplex transition system $S'$ (from $u$ to $v$) such that:

\begin{enumerate}[i)]
\vspace{-\medskipamount}

\item $\pi_{\mathcal P\setminus u,v}(S') \equiv  \pi_{\mathcal P\setminus
    u,v}(S)$,
\vspace{-\medskipamount}

\item $\pi_{u}(S')$ and $\pi_{v}(S')$ are defined over the same sets of messages than  $\pi_{u}(S)$ and $\pi_{v}(S)$.

\vspace{-\medskipamount}

\item $C_{S',u,v} >0$
\end{enumerate}

\end{definition}

%Let us comment definition~\ref{def_CC}. 
The system $S'$ is obtained by replacing processes $u$ and $v$ in $S$ by corrupted processes $u$' and $v$' that implement a covert channel. The half-duplex requirement ensures that a capacity for a supposed covert flow from $u$' to $v$' can be effectively computed. Item $i)$ means that the corrupted system must be equivalent from the non-corrupted users point of view. This means in particular that only processes $u$ and $v$ can be changed. Item $ii)$ means that processes $u$' and $v$' must not implement direct communications from $u$' to $v$', which would be detected by the mechanisms enforcing the security policy. In addition to this, when direct and uncensored communications from $u$ to $v$ exist, establishing a covert channel makes no sense. The last item $iii)$ means that $u$' and $v$' implement a channel with non null capacity, i.e a covert channel. 

One may notice that there exists an infinite number of candidate processes to replace $u$ and $v$ while satisfying conditions $i)$ and $ii)$. We can however restrict arbitrarily our search to systems that are equivalent up to a bounded size. One can immediately remark that when a system $S$ is seen as the composition of independent processes that communicate via channels, i.e. $S=S_u||S_1|| \dots || S_k|| S_v$, $S'$ is obtained by replacing $S_u$ and $S_v$ by some variants $S_u'$ and $S_v'$ that compose similarly with the rest of the system (the sequences of observed messages transiting between $u$,$v$ and the rest of the system are similar). It might be useless to test {\bf all} models up to a certain size, and we think we can limit the search to a finite set of canonical models in which processes $u$ and $v$ have the same number of states but more transitions, and behave as expected by the rest of the system. This remains however to be demonstrated.

% Using a capacity measure to characterize a covert flow allows to consider legitimate channels, i.e. communications that  are hidden in legal information flows. The following definition is close to def.\ref{def_CC}, but quantifies an increment in the capacity of a channel between two users.
%
%INSERT *9
%
Definition~\ref{def_CC} characterizes covert flows in a situation where communications from $u$ to $v$ are forbidden by the security policy. However, covert flows can also appear over legal communications (this is for instance the case of TCP piggybacking). In such situations, a security policy may consist in monitoring or record all messages from $u$ to $v$, and forbid illegal message or contents. Covert flow in this context are called {\em legitimate channels} and their purpose is to bypass the monitoring mechanism.The example of section~\ref{section_example} should be considered as a legitimate channel, as collectors and central observer are allowed to communicate.

\vspace{-\medskipamount}
\begin{definition}
\label{def_CC_legitimate}
A transition system $S$ contains a  {\em legitimate covert channel} from $u$ to $v$ if $u$ and $v$ are allowed to communicate by the security policy, and there exists a Half duplex transition system $S$' such that $\pi_{u}(S')$ and $\pi_{v}(S')$ are defined over the same alphabets as $\pi_{u}(S)$ et $\pi_{v}(S)$, $\pi_{\mathcal P\setminus u,v}(S') \equiv  \pi_{\mathcal P\setminus
    u,v}(S)$, and $C_{S',u,v} - C_{S,u,v}>0$
\end{definition}
\vspace{-\medskipamount}

Definition~\ref{def_CC_legitimate} characterizes situations where modification of the behavior of two processes can add information to the legal contents exchanged when using the original system. Coming back to the example of section~\ref{section_example}, we can notice that there exists no information flow from $CT$ to $CO$ in the original model, even if both processes are allowed to exchange acknowledgment messages. The communication state channel computed from this description alternates channels where only a single input/output is allowed (one state allows to send $ack$, the other allows to send $nack$), hence with zero capacity. This is not surprising, as in the original specification $CT$ only forwards to $CO$ a choice from the environment that is independent from its own actions. Note however that the initial capacity of information flows between two processes that can establish a covert channel is not necessary null. Indeed, some communications can be allowed by the security policy. A covert channel should then be seen as a mechanism that increases the capacity of information flows between two designated parties (the designated sender and receiver processes).

So far, definitions~\ref{def_CC} and~\ref{def_CC_legitimate} remain imprecise on the security policy. Both definitions could require that $S'$ complies with the security policy, but this means expressing this policy as a formal model. For instance, if the security policy is enforced by a monitoring process $P_M$ that detects deviations from a normal behavior, this additional requirement is ensured by  $ii)~~\pi_{\mathcal P\setminus u,v}(S') \equiv  \pi_{\mathcal P\setminus u,v}(S)$. In the global security policy is given as a transition system $S_{sec}$, we simply must ensure that  $\mathcal L(S')\subseteq \mathcal L(S_{sec})$.

\vspace{-3\medskipamount}
\section{Conclusion}
\label{section_conclusion}
\vspace{-\medskipamount}

We have shown how to characterize some covert channels using transition systems and information theory. This characterization shows that a chosen pair of users can simulate a communication channel.  We then translate this transition system to a finite state channel model, for which 
we have effective means to compute a capacity. A capacity greater than $0$ means that the chosen users can establish a covert flow. This characterization works even for covert information flows with capacity lower than 1 bit, but is restricted to a class of transition systems called half-duplex systems. This restriction is mainly motivated by the obligation to compute a capacity for information flows. It imposes in particular that all inputs to the covert channels are independent from the past. However, in many systems, an input and the corresponding outputs can influence the next state. Our 
techniques should be extended to handle this situation. Note however that effective capacity approximation for channels with memory it is still an open question in information theory. We do not expect to obtain closed forms for capacities of covert flows, but good upper bounds can certainly be achieved, and are still useful to characterize leaks. 
We also have assumed that the output alphabet in a covert channel was a code. This assumption was mainly written to simplify the description of our translation from transition systems to communication channels. It can be easily removed at the cost of an approximation, as uncertainty in factorization of received outputs can be seen as adding some randomness in a channel. Similarly, our channel model supposes that the sender has perfect information on the state of the system. This is not always the case in real systems, but capacities achieved with imperfect information are necessarily lower than with perfect information, so this assumption causes no harm to the proposed characterization.

Our covert channel characterization defines a covert flow as the possibility to corrupt a system in an unobservable way. This definition does not bring an effective algorithm to check for covert channels, as there might be an infinite number of such variants of a system. Current solution is to bound the memory of attackers, and work with variant models of size up to this limit. We think however that the search can be limited to a set of canonical variants, as creating new states in the corrupted processes simply means unfolding the original processes in a way that is equivalent to the original observation. In a memoryless system, this does not seem to bring any more encoding power to attackers than allowing more behaviors with the same number of states. This however has to be demonstrated. 

%
%INSERT *10
%
Last, notice that the proposed characterization 
only deals with a specific kind of covert flow, and is defined for transition systems. It  does not consider time, and exhibits a memoryless communication channel with state between two users. We do not know yet if this approach generalizes to wider classes of models, covert flows, and communication channels. Of course, we do not expect to be able to characterize any kind of covert flow using this kind of technique, and the approach proposed in this paper should be seen as complementary of other security tools such as non-interference. However, we think that such characterization captures the essence of covert channels, that is the capacity for a pair of users to establish a communication. 

% \section*{Acknowledgements}
% Many thanks to our partners in the DOTS project, and especially to Béatrice Bérard for careful reading of a preliminary version of this work.

%\nocite{*}
%\vspace{-1.5\bigskipamount}
%\bibliographystyle{plain}
\bibliographystyle{eptcs}

%\bibliography{Coverts}

\end{document}